\documentclass[twocolumn,showpacs,preprintnumbers,amsmath,amssymb]{revtex4}

\usepackage[dvipdf]{graphicx}
\usepackage{dcolumn}
\usepackage{bm}

\begin{document}


\title{Calculation of the Phase Field of a Vortex Pair \\ on the Surface of a Multiplly Connected Substrate }

\author{Toshiaki Obata and Minoru Kubota}
\affiliation{
The Institute for Solid State Physics, University of Tokyo, 
5-1-5 Kashiwa, Kashiwa, Chiba 277-8581, Japan
}

\date{\today}

\begin{abstract}
The vortex pair phase field is calculated, through a solution of the Laplace equation, for a superfluid film 
adsorbed on the surface of a model 3D connected porous medium.  
A number of different vortex-antivortex pair configurations are considered as functions of the porosity 
or an aspect ratio of the model porous structure.  The calculated results give 
the energy of a pair for motion of the vortex cores around a pore or a solid rod as well as for the \textit{string} 
configuration corresponding to a linear separation of the vortex pair through the porous medium.  
\end{abstract}


\maketitle

The superfluid properties of thin $^4 {\textrm He}$ films adsorbed on 3-D connected porous media have been studied 
for many years.  These systems present a number of interesting physical phenomena: dilute bose gas 
behavior for low density $^4 {\textrm He}$ adsorbed in Vycor glass \cite{Reppy00}; 3-D critical behavior observed 
for a wide range of systems\cite{Reppy77}; and 2-D and 3-D crossover behavior as a function of system structure and 
pore size.  
Helium films adsorbed on large-scale flat substrates are known to exhibit the classic 2-D Kosterlitz-Thouless 
phase transition \cite{KT73, Reppy78, AHNS}.  In the case of $^4 {\textrm He}$ films adsorbed on the 3-D 
connected substrate provided by a porous medium, it is clear that one is not dealing with ideal 2-D system; 
yet there remains the interesting question as to what extent vestigial aspects of 2-D K-T behavior may 
still be present in the superfluid behavior of the system \cite{Shira90}.  This question has attracted the attention of 
a number of theorists, the first being Minoguchi and Nagaoka (MN) \cite{Mino88} who argued for the existence 
of a 3-D superfluid phase transition for 3-D connected films and discussed vortex dynamics and the interrelation 
between 2-D and 3-D superfluid transitions.  
These authors pointed out that the energy of a vortex antivortex pair would have a \textit{string-like} 
character with the energy growing linearly for large vortex separations.  
Subsequently, Machta and Guyer (MG) \cite{Guyer88, Machta89} considered 
the 3-D porous media problem as well as the case of a film on a cylinder with infinite length.  
These authors, MN and MG, arrived at two main conclusions.  
First, that vortex pairs are confined by the \textit{string} interaction and consequently a true vortex unbinding transition
of the 2-D K-T type does not occur in a 3-D connected porous medium; and second, 
that the 3-D connectivity of the system enforces 3-D X-Y critical behavior at the superfluid transition.  
Gallet and Williams \cite{Wil89} were the first to emphasis that the \textit{string} 
created by two vortices moving apart in the porous medium will not only create circulation around 
the solid surface over which the pair separates, but also around any loops through which a member of the pair moves.  
In all of the above papers the real complexity of the 3-D connected system is avoided through simplifying assumptions.  

\begin{figure}
\includegraphics[scale=0.4]{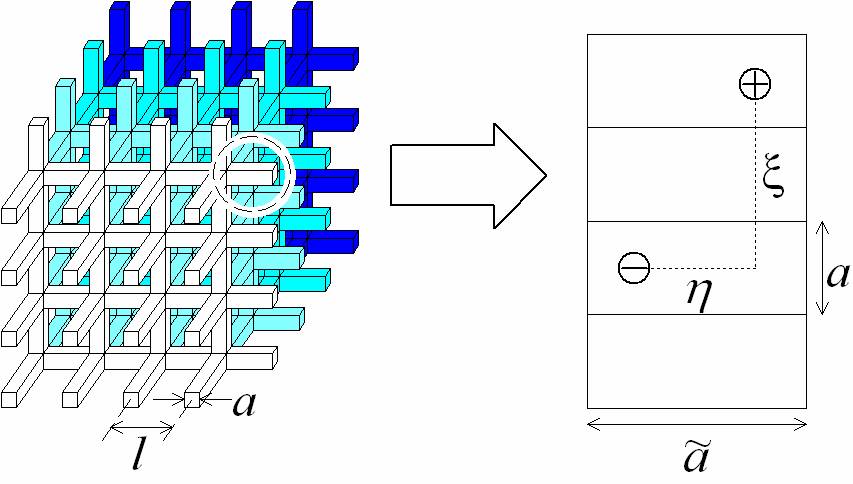}
\caption{\label{fig:porous_cordinate1} The porous substrate and the development of a rod.  
Every rod has the same dimensions.  
The lattice parameter and the width of the rod are defined in the figure.  
$\tilde a = 2 r_0$ and $a = \pi r_0/2$ for comparisons with the cylinder model \protect\cite{Machta89}.  
Then the porosity is 59\%.  
The entire system has a translational symmetry in each three directions.  }
\end{figure}

In the present work we extend the work of the previous authors with a more realistic calculation 
of the energies for quantized vortex pairs taking into account the details of the 3-D porous structure.  
Following MN and MG, we employ the ``Jungle Gym'' model for the porous medium.  
The original model consisted of a cubic structure composed of cylindrical rods with radius $r_0$.  
In our case, we shall construct the lattice with rods of square cross section.  

Figure \ref{fig:porous_cordinate1} shows a view of our lattice structure.  
An aspect ratio for this structure is given by, 
$
\gamma  = {l \mathord{\left/
 {\vphantom {l a}} \right.
 \kern-\nulldelimiterspace} a}
$
where $l$ is the lattice parameter (distance between nodes) and $a$ is the width of the square rods.  
The ratio of the open pore volume to the total volume for a unit cell of the lattice defines the porosity, 
$
P = 1 - {{\left( {3\gamma  - 2} \right)} \mathord{\left/
 {\vphantom {{\left( {3\gamma  - 2} \right)} {\gamma ^3 }}} \right. 
 \kern-\nulldelimiterspace} {\gamma ^3 }}
$.  
The open volume of the pores forms a complementary cubic structure with the same lattice parameter and an aspect ratio, 
$
\tilde \gamma  = {l \mathord{\left/
 {\vphantom {l {\tilde a}}} \right.
 \kern-\nulldelimiterspace} {\tilde a}} = {\gamma  \mathord{\left/
 {\vphantom {\gamma  {\left( {\gamma  - 1} \right)}}} \right.
 \kern-\nulldelimiterspace} {\left( {\gamma  - 1} \right)}}
$, where $\tilde a = l - a$, is the length of a rod or the edge length of a pore.  In making a comparison to 
the MG model, we shall initially choose a lattice parameter, $
l = \left( {2 + {\pi  \mathord{\left/
 {\vphantom {\pi  2}} \right.
 \kern-\nulldelimiterspace} 2}} \right)r_0 
$ and set $
a = {{\pi r_0 } \mathord{\left/
 {\vphantom {{\pi r_0 } 2}} \right.
 \kern-\nulldelimiterspace} 2}
$.  With this choice the length of the rods will be, $\tilde a = 2 r_0$ and the porosity has a value close to 0.59.  

In the present calculation the vortex pair phase field is computed on a 4$\times $4$\times $4 cubic lattice 
with 64 surface and interior lattice nodes and 192 rods.  Periodic boundary conditions can be imposed 
that insure that the gradient in the phase normal to the surface of the cube is zero, i.e., the flow field induced 
by the vortex pair is restricted to the interior of the cube.  A more ambitious project would be to perform 
these calculations on a larger lattice; for instance, one might take a 6$\times $6$\times $6 cubic structure 
with 216 nodes and 648 rods to be considered in the calculation.  

We shall assume that a uniform superfluid film with density per unit area, 
$\rho$, covers the surface of the lattice.  
In the present calculation, we will take into account the actual structure of the lattice 
in calculating the phase and superfluid flow fields for various configurations of a vortex-antivortex pair 
located on the surface of the lattice.  For simplicity we start with the cores located close together on 
a single lattice rod.  This rod will be chosen as close to the center of the lattice as possible given 
the discrete nature of the structure.  The surface of the rod can be unfolded as illustrated in Figure \ref{fig:porous_cordinate1}.  
A coordinate $\eta$ indicates the distance between the vortex cores along the axis of the rod while
$\xi$ gives the distance in the perpendicular direction (along the perimeter of the rod).  
In the figure the vortex and antivortex cores are shown placed symmetrically about the center of the rod. 
The rod containing the vortex pair has a common boundary with four other adjacent rods and the phase field must be 
matched continuously across the common edges.  

At the start of the calculation, the initial position of the vortex pair is specified to be close together on an interior rod.  
The quantum phase, $\phi$, is calculated as a function of position on the surface of the lattice 
by a numerical solution of Laplace's equation on 64$\times $64 grid on each rods while maintaining phase continuity 
across the matching edges of adjacent rods.  
Once the phase field is known, the superfluid velocity is obtained from 
the relation, $ {\bf v}({\bf r}) = (\hbar /m_4 )\nabla \phi ({\bf r}) $
 and the total energy from an integration of $(1/2)\rho {\bf v}^2 $
 over the surface of the lattice.  

\begin{figure}

\begin{center}
(a)
\includegraphics[scale=0.3]{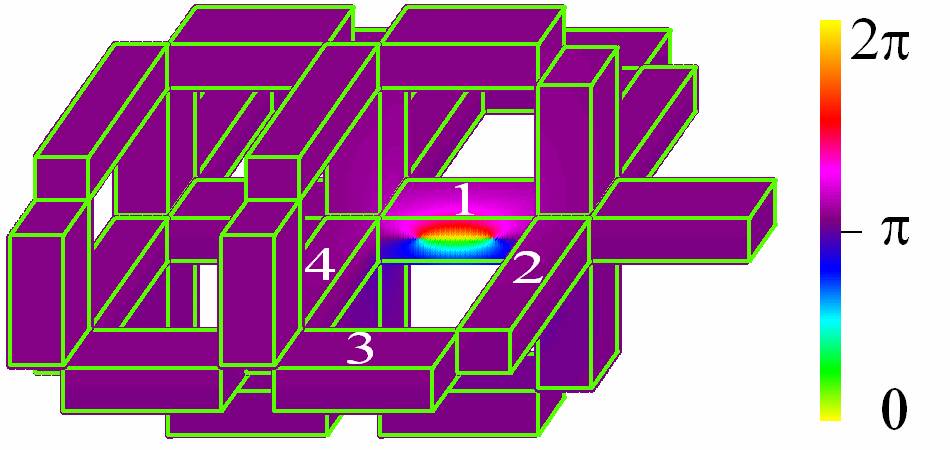} 
\end{center}

\begin{center}
(b)
\includegraphics[scale=0.3]{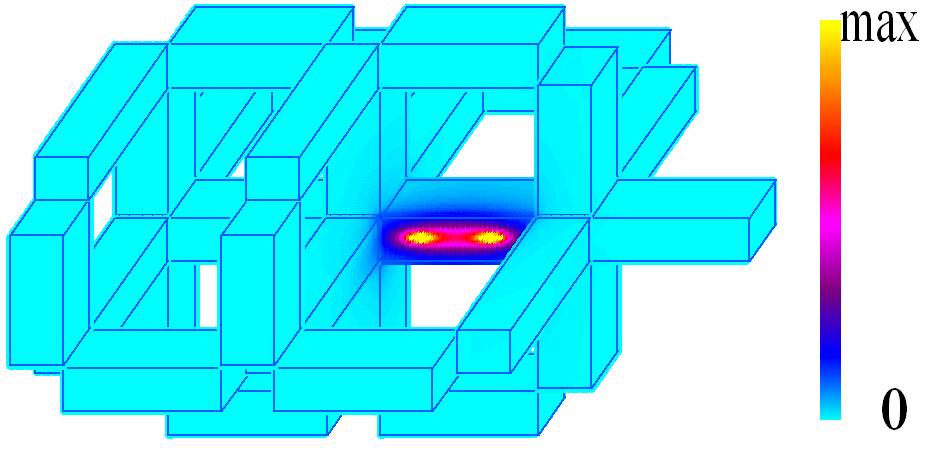} 
\end{center}

\begin{center}
(c)
\includegraphics[scale=0.3]{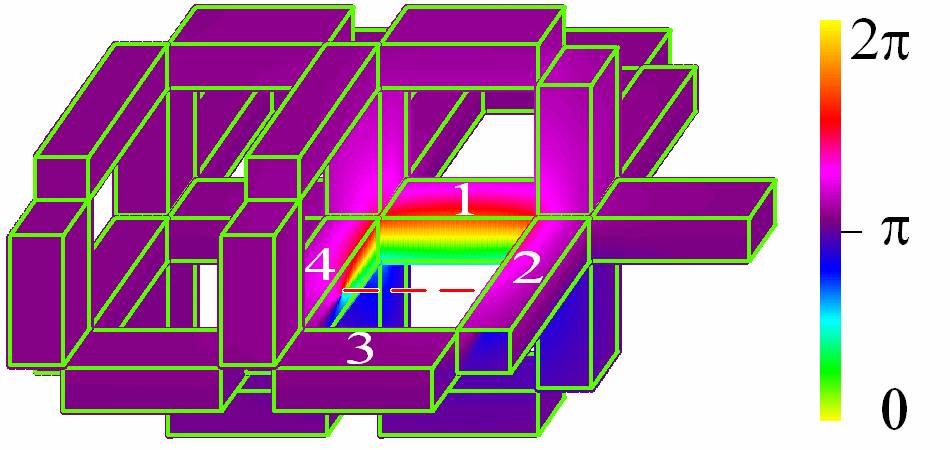}
\end{center}

\begin{center}
(d)
\includegraphics[scale=0.3]{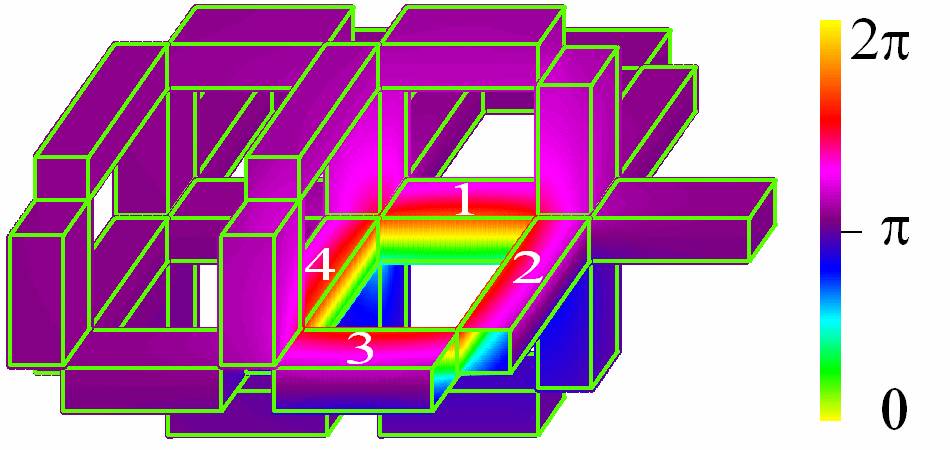}
\end{center}

\caption{\label{fig:vp_eta} A series of the calculation results.  This solution is
the example when vortex pair decouples along $\eta$ direction.  
The color corresponds to the phase from 0 to 2$\pi$ and it is periodical continuous in the panels a, c and d.  
(a) $\eta = \tilde a / 2 $.  
(b) The flow energy distribution of the configuration (a).  
(c) $\eta = 2 \tilde a $.  The red broken line indicates the vortex core crossing the pore.  
(d) The symmetric flow pattern remaining after vortex pair annihilation.  }
\end{figure}

In Figure \ref{fig:vp_eta}, three different configurations for a vortex pair are shown.  
In the first panel, \ref{fig:vp_eta}a, the vortices are located on the rod labeled 1 and are separated by a distance,
$\eta$ = $\tilde a / 2$.  Phase values between 0 and $2\pi$ for this pair configuration are indicated by the color scheme.  
In panel \ref{fig:vp_eta}b the color scale indicates the superfluid energy distribution.  
When the vortex pair is restricted to a single rod there is very little flow on the adjacent rods.  
When the vortices are moved futher apart, until they are located at the corner intersection of the rods, 
the vortex flow field spills over onto the surfaces of the adjacent rods.  
In \ref{fig:vp_eta}c the vortices are shown at the mid point of rods 2 and 4.  
The vortices now face each other across the pore.  As the vortices are moved apart, it is useful to keep track of 
the virtual core associated with the vortex-antivortex pair.  The core will form a closed loop intersecting the surface 
of the rods at the position of the vortices.  The core then has two parts; one within the interior of the solid rods 
and another portion within the space of the open pores.  
When the separation is further increased to $\eta  = 4 \tilde a$
, the two elements of the pair will then overlap and the virtual core of the vortices will lie entirely within the rods.  
We are now left with a ``core'' vortex configuration, with a four-fold symmetric flow largely restricted 
to the circumference of the rods, 1 to 4.  We shall call this flow configuration the $\eta$ - ring 
and describe the phase configuration in Figure \ref{fig:vp_eta}d.  

\begin{figure}
\begin{center}
(a)
\includegraphics[scale=0.3]{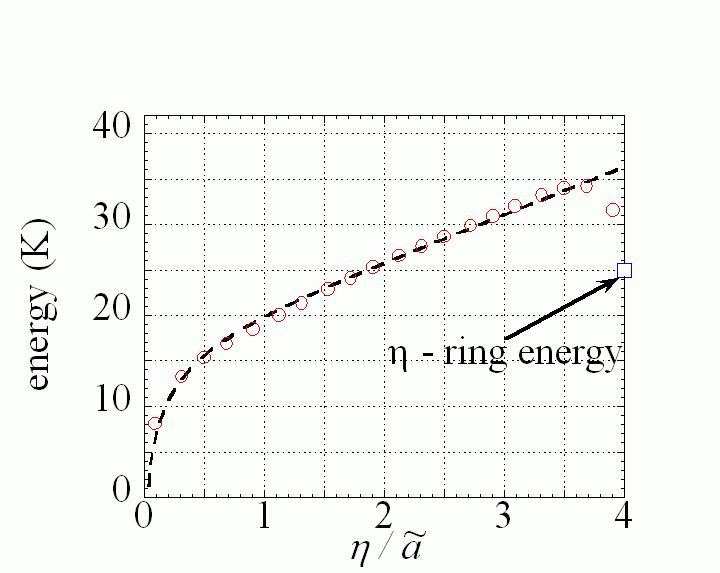}
\end{center}
\begin{center}
(b)
\includegraphics[scale=0.3]{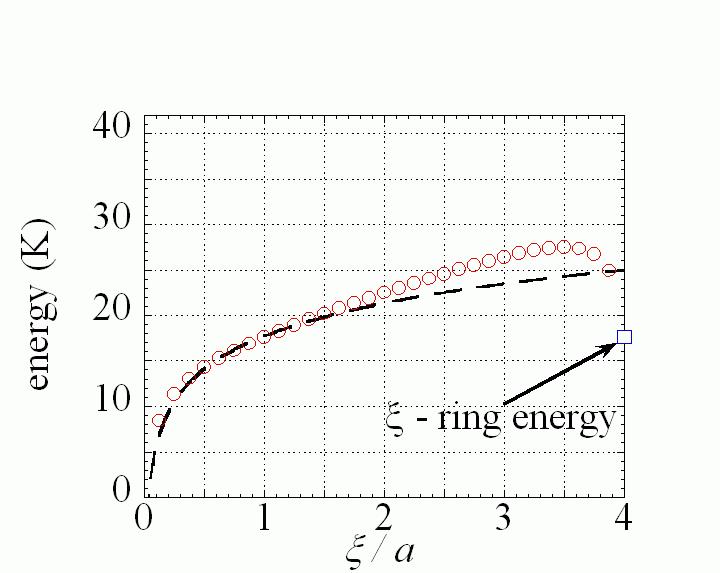}
\end{center}
\begin{center}
(c)
\includegraphics[scale=0.3]{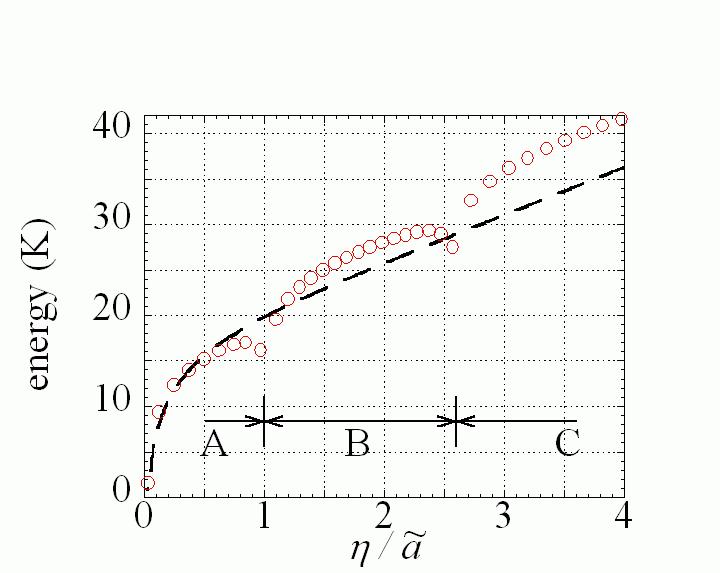}
\end{center}

\caption{\label{fig:eta_dep} 
The vortex pair energy functions.  \\
(a) The energy function when a vortex pair expands 
along the $\eta$ axis.  
The termination points are 
the annihilation points where a vortex pair annihilates.  
(b) The energy function when a vortex pair expands 
along the $\xi$ axis.  The broken line is a normal K-T type logarithmic function.  
(c) The case when a vortex pair expand along the $\eta$ direction through a pore \protect\cite{Wil89}.  
The label A, B and C corresponds to the indicators in Fig. \ref{fig:override}.   \\
The squares in the panels a and b are the $\eta$ and $\xi$ ring energy respectively (see the main text). 
The broken lines in panels a and c are the analytic cylinder function \protect\cite{Machta89}. 
}
\end{figure}

We now examine the change in the kinetic energy of the flow field as the pair separates along $\eta$ direction.  
In Figure \ref{fig:eta_dep}a we show a plot of the total energy as a function of the vortex-antivortex 
separation distance $\eta$.  
In setting the energy scale, a superfluid mass per unit area is taken to be, $\rho = 3.49 \times 10 ^{-9} gm / cm^2$, 
so that a K-T transition causes at 1K in the case of the infinitely spreaded $^4 He$ film.  
The infinite cylinder calculation of MG\cite{Machta89} is shown for comparison.  
It is interesting that these two calculations agree rather well even as the vortex pair passes a corner 
in the lattice structure.  

\begin{figure}
\begin{center}
(a)
\includegraphics[scale=0.3]{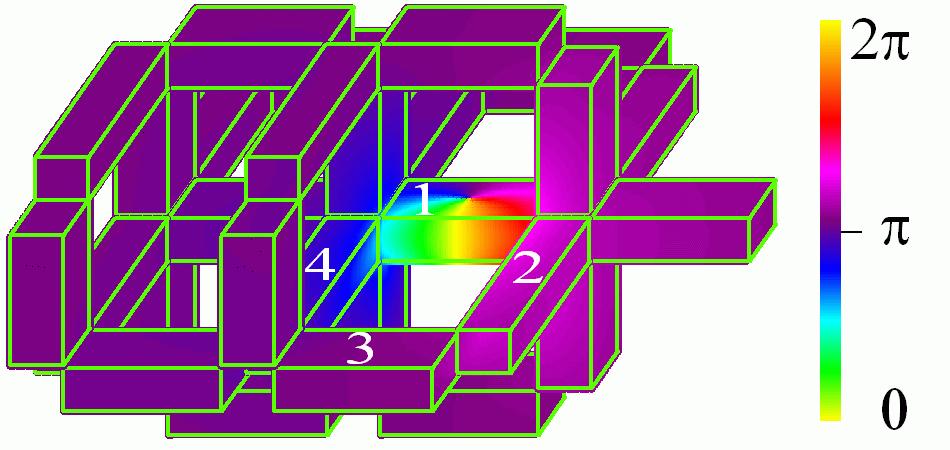}
\end{center}
\begin{center}
(b)
\includegraphics[scale=0.3]{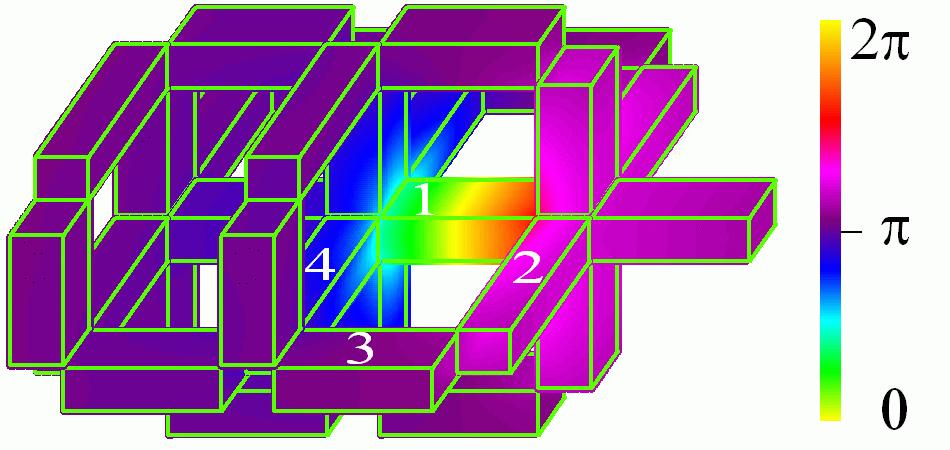}
\end{center}
\caption{\label{fig:vp_xi} The calculation results when the vortex pair expands 
in the $\xi$ direction.  (a) $\xi = \pi r_0$.  
(b) The case of a pore vortex ring (see the main text.  ).  }
\end{figure}

We shall now consider pair separation in the $\xi$ direction, perpendicular to $\eta$, 
along the perimeter direction of the rod.  The initial configuration is taken with the pair located 
at the center of the rod ($\eta= 0$) separated by a short distance in the $\xi$ direction 
and the vortex pair moves on in the $\xi$ direction.  
When the separation is the perimeter distance, $\xi = 2 \pi r_0$, the pair will have encircled the rod and the cores 
will overlap and leave behind a continuous flow field along the axis of the rod.  
Figure \ref{fig:vp_xi} shows the phase field on the surface of the lattice as the vortex pair expands in the 
$\xi$ direction.  
The development of the energy as a function of the separation parameter, $\xi$, is shown in panel b of Figure \ref{fig:eta_dep}.  
The logarithmic dependence of the pair energy for a vortex-antivortex pair in 2-D is 
also shown as a function of pair separation.  An interesting feature of the energy dependence is that 
a maximum occurs in the total energy shortly before the pair annihilates on the far side of the rod.  
After the annihilation, the system is left in a symmetric flow state similar to the flow pattern of a vortex ring in 3-D space; 
in the present case, however, the flow is restricted to the surface of the lattice.  
This remaining flow configuration had been called the ``pore'' 
vortex ring structure with the virtual vortex core threading the pores immediately surrounding the rod\cite{Guyer88}.  
Here, we call this type of a pore vortex ring the $\xi$ - ring.  

\begin{figure}
\begin{center}
(a)
\includegraphics[scale=0.3]{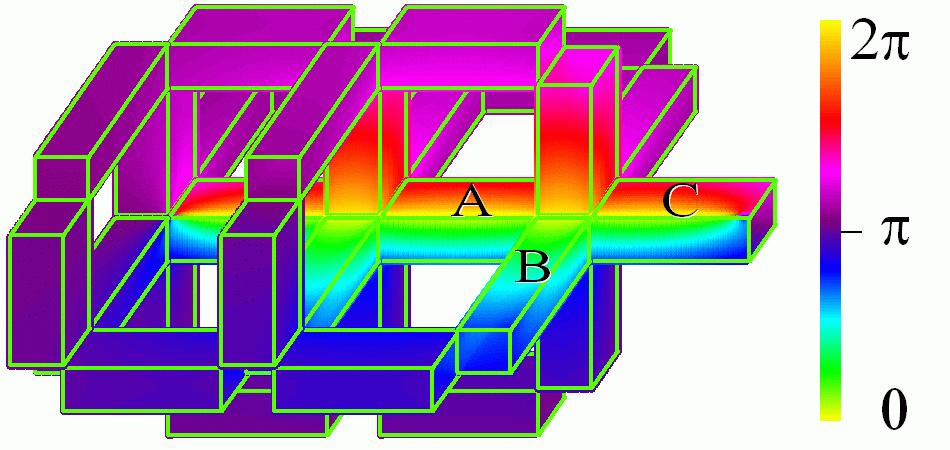}
\end{center}
\begin{center}
(b)
\includegraphics[scale=0.3]{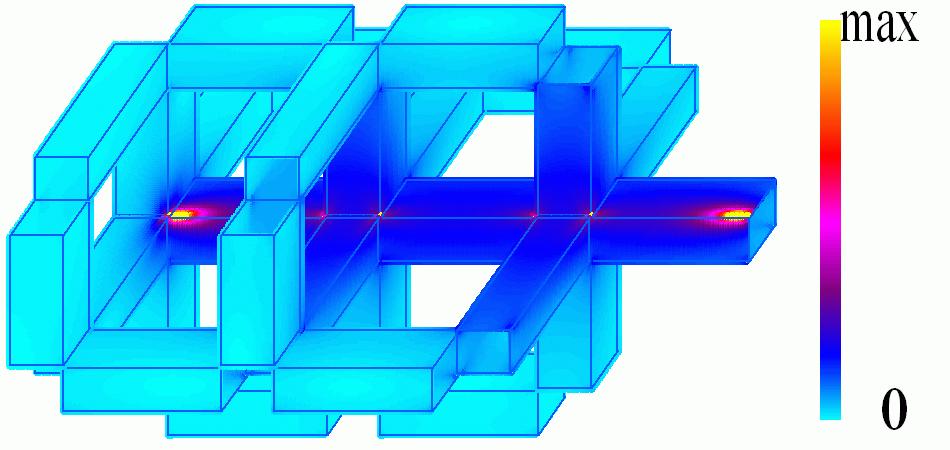}
\end{center}

\caption{\label{fig:override} 
A calculation result that vortex pair expands through pores.  
A vortex pair symmetrically expands from the middle point 
of an edge of the rod A.  (a) The phase is described in color scheme.  
The indicators A, B and C correspond to the indicators in Figure \ref{fig:eta_dep}.  
(b) The energy distribution of the same arrangement.  At each corner where vortex have gone through, there are small 
energy concentration points.  }
\end{figure}

A third configuration, shown in Figure \ref{fig:override}, deserves consideration.  
The starting configuration is similar to the first case considered, with the vortices displaced 
in the $\eta$ direction.  We shall take the initial position of the vortices 
on the edge of the rod rather than in the middle of one face.  
Now, as $\eta$ increases, 
the vortices can move in a straight line along the junction between rods and so on to the rods beyond.  
When a pair is separated with the elements moving past a junction in the lattice, 
from one rod to another, circulation is created not only around rods which have virtual core within 
but also about any loop in the lattice threaded by a single line of the core.  
Thus the energy of a separating vortex pair does not only grow linearly with separation in the \textit{string} fashion 
but increases in a step-like fashion whenever a vortex moves through a pore.  The result is an even stronger ``confinement'' \cite{Mino88}
than results from the simple \textit{string} interaction.  

\begin{figure}
\includegraphics[scale=0.3]{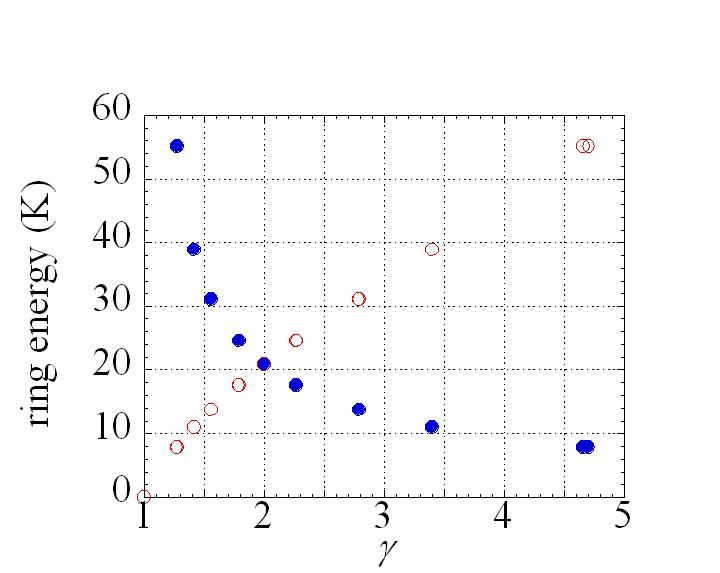}
\caption{\label{fig:poro_E} The $\eta$ and $\xi$ ring energies are shown as open and closed circles, respectively,
as functions of the aspect ratio $\gamma$.  
}
\end{figure}

We have also calculated the ring energies for different values of $\gamma$ (or porosoity).  
The results are shown in Figure \ref{fig:poro_E}.  
E$_{ring} ^{\xi}$ is the $\xi$ - ring energy 
while E$_{ring} ^{\eta}$ the $\eta$ - ring energy.
From symmetrical point of view, E$_{ring} ^{\eta}$ for 30\% porosity is E$_{ring} ^{\xi}$
for 70\% porosity and vice versa; in other words E$_{ring} ^{\eta}$ for a certain $\gamma$ is E$_{ring} ^{\xi}$ for 
the same $\tilde \gamma$ value as $\gamma$.  
For small values of $\gamma$, the $\eta$ - ring energy is nearly a logarithmic function of $\gamma$, 
while for large $\gamma$, it approaches a nearly linear dependence of $\gamma$.  
$\gamma = 1$ corresponds to the case that $l = a$; namely, there is no pore then no film let in the system.  
For comparison with this energy dependence it is worth considering a constant flow 
just on the surface of four adjacent rods forming a pore.  
We shall assume the magnitude of the velocity is fed by the phase difference of $2 \pi $; i.e. $v = h / 4 a m_4$.  
With this velocity, the total energy E 
$= 4 \times \left[ {(\rho /2)v^2 } \right] \times 4a\tilde a = 12.6 ( \gamma - 1 )$ (K), where we have again set 
$\rho = 3.49 \times 10^{-9} gm / cm^2$ in setting the energy scale.  
This energy function is a little smaller than E$_{ring} ^{\eta}$ but parallel for large values of $\gamma$.  
This difference is caused by the fact that the actual velocity distribution is not constant because of the 3D connectivity.  

The energy dependence tells us whether a pore vortex ring could be excited or not.  
Around $\gamma = 2$, both E$_{ring} ^{\eta}$ and E$_{ring} ^{\xi}$ are too high to be excited 
and a pore vortex ring would be rarely excited.  Because we now know the energy behaves like a cylinder case 
in both $\xi$ and $\eta$ direction until it approaches the maximum, we could constract the renomalization calculation 
by an imitation energy function, which expresses a vortex string energy for both axes.  
The study would be closely connected to the experimental study on 
the Vycor glass substrate \cite{Reppy77} as well as on the other Porous glass substrates \cite{Shira90}.  
As $\gamma$ increases, the porosity becomes higher and a pore vortex ring comes to be excited more easily.  
Then some simple mean field theory would better describe the physical property 
of the system like $^4He$ film on aero gel \cite{Crowell95}.  

The authors thank many people.  We thank Prof. J.D. Reppy for fruitful discussions and suggestions.  
The colored graph was described using 
the basic graphics class library for C by Mr. K. Yamada\cite{graphicclass}.  

\bibliography{porous_cal}

\end{document}